# Modified Band Alignment Method to Obtain Hybrid Functional Accuracy from Standard DFT: Application to Defects in Highly Mismatched III-V:Bi Alloys.

Maciej P. Polak,[1,2,*] Robert Kudrawiec,[2] Ryan Jacobs,[1] Izabela Szlufarska,[1] and Dane Morgan[1,†]

[1] *Department of Materials Science and Engineering, University of Wisconsin-Madison, Madison, Wisconsin 53706-1595, USA*

[2] *Department of Semiconductor Materials Engineering, Faculty of Fundamental Problems of Technology, Wrocław University of Science and Technology, Wybrzeże Wyspiańskiego 27, 50-370 Wrocław, Poland*

This paper provides an accurate theoretical defect energy database for pure and Bi-containing III-V (III-V:Bi) materials and investigates efficient methods for high-throughput defect calculations based on corrections of results obtained with local and semi-local functionals. Point defects as well as nearest-neighbor and second-nearest-neighbor pair defects were investigated in charge states ranging from -5 to 5. Ga-V:Bi systems (GaP:Bi, GaAs:Bi, and GaSb:Bi) were thoroughly investigated with significantly slower, higher fidelity hybrid Heyd-Scuseria-Ernzerhof (HSE) and significantly faster, lower fidelity local density approximation (LDA) calculations. In both approaches spurious electrostatic interactions were corrected with the Freysoldt correction. The results were verified against available experimental results and used to assess the accuracy of a previous band alignment correction. Here, a modified band alignment method is proposed in order to better predict the HSE values from the LDA ones. The proposed method allows prediction of defect energies with values that approximate those from the HSE functional at the computational cost of LDA (about 20× faster for the systems studied here). Tests of selected point defects in In-V:Bi materials resulted in corrected LDA values having a mean absolute error (MAE) = 0.175 eV for defect levels vs. HSE. The method was further verified on an external database of defects and impurities in CdX (X=S, Se, Te) systems, yielding a MAE = 0.194 eV. These tests demonstrate the correction to be sufficient for qualitative and semi-quantitative predictions, and may suggest transferability to many semiconductor systems without significant loss in accuracy. Properties of the remaining In-V:Bi defects and all Al-V:Bi defects were predicted with the use of the modified band alignment method.

## I. INTRODUCTION

There are many challenges to obtaining defect properties from direct experimental characterization, therefore theoretical predictions of point defect properties are very valuable. For example, DLTS is widely used to study defects (traps for carriers) in semiconductors since this technique provides experimental information on the trap energy level, capture cross section and trap concentration but lacks the ability to identify the defect type [1]. Advanced electron microscopy techniques are able to identify the defect type [2–4] but the measurements are expensive, destructive, difficult to properly perform, and require time-consuming sample preparation.

First principle density functional theory (DFT) calculations of defect levels aid interpretation of results obtained with DLTS or other methods and are often used in defect characterization. DFT also enables exploration of a full suite of defect types and interactions and offer physical insight into defect and associated electronic and optical properties of a material. Correct DFT treatment of defects in semiconductors requires the use of supercells with many atoms and a proper description of both the total energy of the system and its electronic band structure. Recent developments of computational methods such as hybrid functionals and corrections of the spurious electrostatic interactions, allow for accurate theoretical predictions. However, the use of these high fidelity methods also results in very time consuming and computationally expensive calculations, which typically limits researchers to focus only on a small number of specific defect types or a particular material system. Therefore, in order to investigate large range of systems, defects and charge states, approximate but more efficient methods are necessary.

The goal of this work is two-fold. First, to provide and assess potential methods to reduce the computational effort required to obtain defect formation energies and defect levels, while preserving relatively high accuracy. Here, band alignment based corrections, as well as machine learning methods, are investigated. Second, to use the most efficient of these methods to build a highly-accurate first principles defect energy database for pure and Bi-containing III-V (III-V:Bi) materials.

Dilute bismides are group III-V semiconductors with bismuth as an isovalent dopant replacing a modest percentage (typically 10% or less) of the group V host atoms. They have attracted interest in the past 10 or so years mostly due to a significant reduction of the band gap with relatively low Bi concentration and a large increase in the spin-orbit splitting [5–8]. Specifically, when the Bi concentration in $GaAs_{1-x}Bi_x$ approaches 12% the spin-orbit splitting exceeds the band gap energy, Auger recombination is suppressed which results in higher device efficiency when the material is used in long wavelength laser devices [9]. Additionally, alloying with bismuth is the only way to obtain a group III-V material with a band gap lower than that of $InAs_{1-x}Sb_x$, which is necessary for mid-infrared (MIR) devices. So far, the MIR spectral range is usually covered by using the group II-VI alloy $Hg_{1-x}Cd_xTe$. However, there are many disadvantages of using this material, such as instability and lack of compositional uniformity [10, 11], as well as the use of environmentally hazardous and highly toxic Hg and Cd elements. A III-V compound with properties favorable for use in MIR devices would provide a possibility to overcome the challenges of using $Hg_{1-x}Cd_xTe$. Such a compound would also enable processing of MIR-compatible materials using highly developed III-V industry technologies, allowing easy integration with existing infrastructure. Some optoelectronic structures based on dilute bismides such as photodetectors [12–14] and recently even laser structures [15, 16], have been developed and shown promising results, encouraging further development. It is evident, that dilute bismides are an active field of research, with focus not only on Ga- and In-based III-V:Bi systems, but now also extending to Al- based, with $AlSb_{1-x}Bi_x$ having been synthesized for the first as recently as last year [17].

* mppolak@wisc.edu
† ddmorgan@wisc.edu

Point defects properties are particularly important to establish for dilute bismide semiconductor systems. This importance arises primarily because many dilute bismides can be regarded as highly mismatched semiconductor alloys (HMAs) due to the significant discrepancy of the electronegativity and size of the Bi atoms compared with the atoms comprising the host III-V compounds. Dilute bismide systems are known for being difficult to manufacture. In order to incorporate Bi atoms into the III-V host, the growth conditions have to be adjusted, typically to a significantly lower growth temperature than that optimal for pure III-V host materials [18–23]. As a consequence, undesired defects often form during III-V:Bi growth, which defects often function as traps for carriers, in turn lowering device efficiency.

Because of poor optical quality of HMAs, post growth annealing is often applied in order to increase the efficiency of luminescence [24–26]. It is believed that for dilute nitrides, the enhancement of luminescence is related to reduction of the concentration of point defects due to annealing [24, 27, 28]. In the case of dilute bismides, post growth annealing has also been performed [29–34], but its role in the improvement of the material quality is still unclear. However, defects are expected to be contributing to the annealing response. Overall, defects are likely to play a key role in these systems and knowledge of their properties may help improve materials performance. Specifically, the defect data included in this study can guide the growth process or annealing conditions and enable better interpretation of the results of defect properties measured by spectroscopic, optical or electrical methods, such as deep-level transient spectroscopy (DLTS), photoluminescence, or temperature-dependent Hall measurements. Although a few studies focused on particular materials (GaAs:Bi and GaSb:Bi) and defect types exist [35–38], a comprehensive study has not been performed, until now. The data and knowledge contained in this work provide powerful tools for understanding the performance limitations and improving the quality of grown structures.

An optimized band alignment correction is proposed, where results of significantly faster, lower-fidelity local density approximation (LDA) (and other local and semi-local functionals) calculations are corrected to approach the accuracy of significantly slower, higher-fidelity hybrid Heyd-Scuseria-Ernzerhof functional (HSE) calculations. Related to the correction, it has been proposed that much of the improvement associated with HSE vs. LDA can be obtained by a simple band alignment (BA) correction to the LDA energy levels [39–45]. Such an approach can potentially produce results that come close to HSE-level accuracy at a fraction of the computational cost (typically more than an order of magnitude reduction in computing time). However, the BA approach has only been tested on a handful of defect types and systems, and therefore has not yet been thoroughly assessed and may not yet be fully optimized. To develop our database and explore a band alignment based approach to improving LDA results, all Ga-V:Bi materials were studied directly with both the lower accuracy LDA method and the higher accuracy HSE approach. We determined formation energies for all charge states for native and Bi-related point defects as well as nearest- and second-nearest-neighbor pair defects. Then, we applied the BA correction method to the LDA results and assessed the accuracy of this correction against HSE. Based on the result of the BA correction, we propose an additional empirical correction, which we call the modified band alignment (MBA) method. The MBA includes a band gap-dependent linear shift, which we fit to the Ga-V:Bi training data. We then verify the improved accuracy of the MBA vs. the BA for a test data set of point defects in In-V:Bi. In addition, to test the transferability of the method to other systems that are not III-V semiconductors, the MBA was tested on a large, recently published database of defect and impurity energies in CdX (X=S, Se, Te) materials [46]. Finally, the MBA is used to predict the properties of the remaining pair defects in In-V:Bi together with all defect properties of Al-V:Bi materials. As a result, a large database of formation energies and charge-state transition levels in III-V:Bi materials is generated.

## II. RESULTS AND DISCUSSION

### A. Validation of first principles calculations

In this section, we validate our first principles calculations of charge-state transition energies by comparing to available experimental and computed data. There is limited data for comparison on defect properties of dilute bismides and pure GaP and GaSb, but a significant amount for GaAs, so we focus on GaAs:Bi. DFT studies of GaAs:Bi by Luo *et al.* [47] using a very similar methodology to this work (see Sec. IV) are in an excellent agreement with the calculations performed here. The difference between our predicted point defect charge-state transition levels and those of Luo *et al.* have a MAE of 0.06 eV. Excellent agreement is also found between our calculated and the available point defect charge-state transition levels from experimental results, with a MAE of 0.04 eV, based on our best interpretation of the experimental data. This comparison is done for the same data and defect type and defect levels used in Luo *et al.*

Calculations performed by Buckeridge *et al.* [48] for point defects in pure GaSb use a similar methodology to that used in this paper and are in very good agreement with our calculations, with differences in defect levels not exceeding 0.1 eV. The authors, similarly to us, find only qualitative agreement with the studies performed by Virkkala *et al.* [49] on GaSb. The quantitative disagreement can be attributed to differences in computational procedures, such as convergence criteria, treatment of spin-orbit interaction, and different approach used for the electrostatic corrections.

The experimental observations of metastability of vacancy related defects in GaSb [50], and their theoretical predictions in GaAs [51], where in certain charge states instead of a single group III vacancy ($v_{III}$), a pair antisite defect and group V vacancy is preferred, are also well reproduced in the current work.

A careful investigation of Fig. 1 f), h), and i) reveals that some defects are stable in a negative (-1) charge state at $E_F = 0$ eV. These defects are group III vacancies (in low band gap materials: GaSb, InAs and InSb) and a $Ga_{Sb}$ antisite.

### B. Insight from first principles calculations

The calculated energies for the Ga-V:Bi systems can be used to better understand the experimental observations associated with III-V:Bi materials. They also allow us to make predictions about possible opportunities and challenges for improving the processing of III-V:Bi materials by mitigating the deleterious effect of defects. In the following we assume that higher stability of defects indicates these defects are present at higher concentrations (relative to less stable defects). This claim is certainly true at equilibrium but may not be true under the nonequilibrium growth conditions used for many of these materials. In particular, III-V:Bi alloys are often grown using molecular beam epitaxy and metalorganic vapor phase epitaxy under conditions that are very far from equilibrium. However, we assume that even in these cases the chemical trends and qualitative features identified below are likely to be preserved.

For the case of the isovalent doping ($Bi_V$), low formation energy as well as a stable zero charge defect state with no charge-state transitions is desirable in terms of stably incorporating Bi to form an alloy. Figs. 1 d)-f) show formation energies for point defects in Ga-V:Bi with chemical potentials corresponding to intermediate growth conditions (group III-rich and group V-rich conditions can be found in supplementary Figs. S2 and S8). The predicted formation energies for $Bi_V$ follow the chemical trend of the mismatch between the group V and Bi atoms, with the highest formation energy for GaP and lowest for GaSb. This result is consistent with previous research on growth of these materials, where the most substantial amounts of Bi can be incorporated in GaSb, less Bi can be incorporated in GaAs, and a high Bi composition in GaP is most difficult to achieve. So far, GaSb:Bi layers and quantum wells with quite good optical properties were reported by Kopaczek *et al.* [52], Yue *et al.* [53] and Delorme *et al.* [54] while reports on optical properties of GaP:Bi are still very limited and not promising [55, 56]. This suggests that high-quality GaP material with Bi is difficult to achieve.

Bi-related defects are very important for the optical quality of dilute bismides. In particular, materials can suffer from undesired antisite $Bi_{III}$ defects. A high formation energy of those antisite defects is therefore preferable in order to manufacture a high-quality alloy. In p-type materials, the predicted $Bi_{Ga}$ defect formation energies are very low, close to zero (see Fig. 1 d)-f)). This is especially apparent under V-rich conditions, and in GaP and GaAs, where the formation energy of $Bi_{III}$ is lower than that of $Bi_V$. The situation improves with the increase of Fermi level, where for n-type materials, $Bi_{Ga}$ always exhibits a higher formation energy than $Bi_V$.

The binding energy of the $Bi_V$ and $Bi_{Ga}$ defects in each system is the lowest of all nearest-neighbor-pairs and exhibits almost no change as a function of Fermi level (Fig. 2 d)-f)). This result suggests that no significant clustering between $Bi_V$ and $Bi_{Ga}$ defects should be expected. The formation energies show that among the three Ga-V materials studied, GaSb:Bi (Fig. 1 f)), the material with the lowest mismatch between Bi and group V atom, exhibits the lowest formation energy for isovalent $Bi_V$ doping and the highest energy for $Bi_{Ga}$ antisite under both III- and V-rich conditions. This suggests that GaSb:Bi might be the most promising for obtaining high-quality (i.e. low defect concentration) materials.

The binding energies show the influence of the mismatch between Bi atoms and the corresponding group V atoms very clearly. GaP:Bi and GaAs:Bi have a noticeable similarity in the shapes of the formation energy curves (Fig. 1) as well as the relative values of the binding energies. Defects in these two materials span a similar range of stable charge states, with similar order of the formation energies between different defects. GaSb:Bi does not share this similarity. However, GaSb is the only material out of the three where the group V atom is larger than the group III atom, which results in different lattice-strain related effects for substitutional defects. GaSb:Bi is also the material with the lowest mismatch, due to the similar sizes and electronegativities of Sb and Bi, resulting in a Bi level with much larger separation from the valence band maximum (VBM) than the GaAs:Bi and GaP:Bi. This large separation, in turn, results in different electronic properties near the band gap edges, with significantly lower electron localization and narrower emission peaks, as well as reduced lattice strain in Bi-related defects. The reduced strain effects also encourage higher amounts of Bi to incorporate into the alloy.

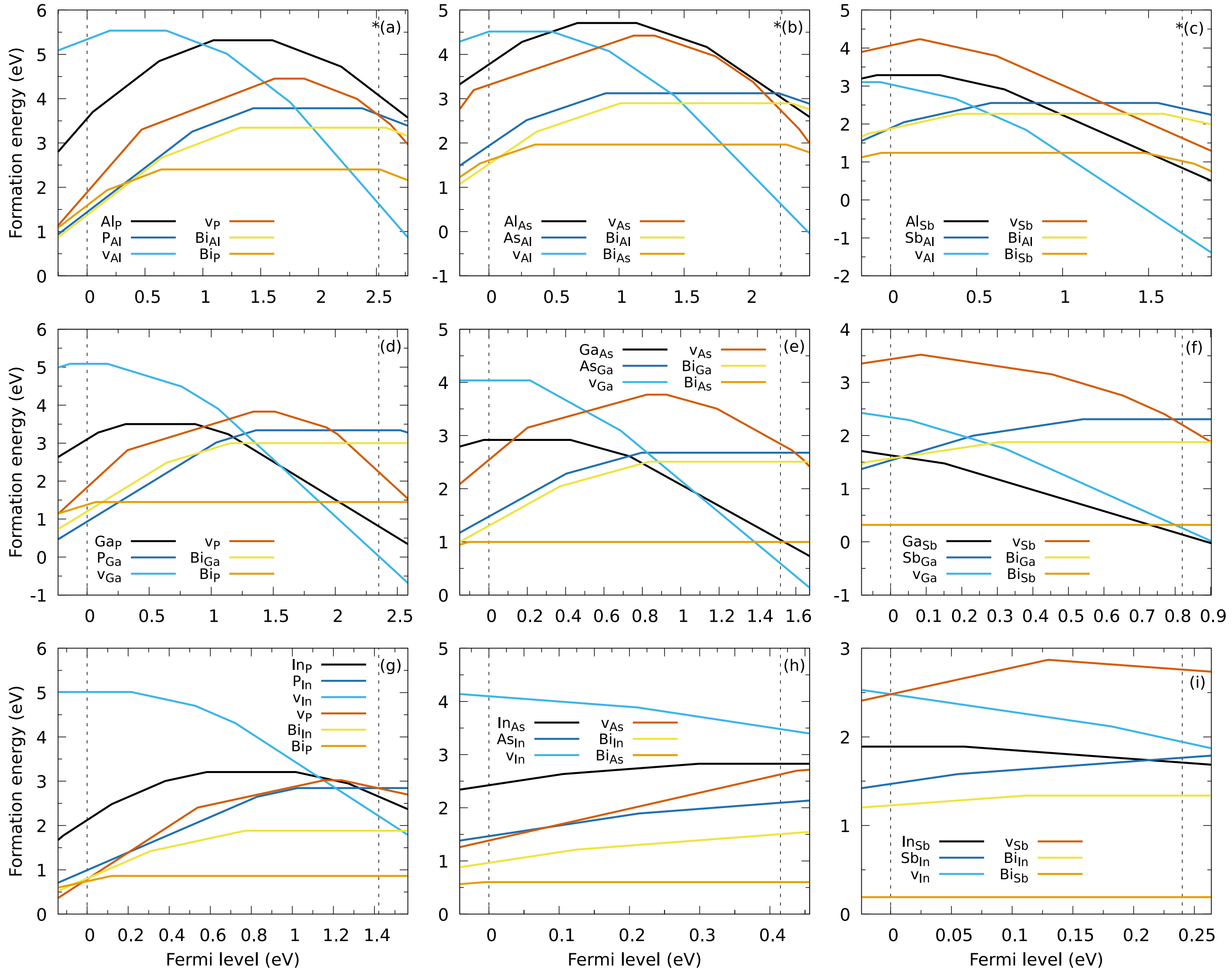

FIG. 1. Point defect formation energies as a function of Fermi level for all considered III-V:Bi materials. Panels d)-i) (Ga-V:Bi and In-V:Bi) were calculated with the HSE functional, and a)-c) (Al-V:Bi) were obtained with LDA corrected with the modified band alignment correction (MBA) and, therefore, are marked with an asterisk. a)-c) correspond to AlP:Bi, AlAs:Bi, and AlSb:Bi, d)-f) to GaP:Bi, GaAs:Bi, and GaSb:Bi, and g)-i) to InP:Bi, InAs:Bi, and InSb:Bi, respectively. Chemical potentials corresponding to intermediate growth conditions were used. Results for group V- and group III-rich conditions are available in Sec. VI A, Figs. S1-S3 and S7-S9. Here we assume the MBA is sufficient for quantitative comparison, therefore, this figure serves as a presentation of the results for point defects in all studied systems, and not for assessing the method, for which we direct the reader to Figs. 3 and 4.

Fig. 2 shows binding energies of pair defects in Ga-V:Bi materials. P-type GaP:Bi and GaAs:Bi have a large (attractive) binding energy for nearest-neighbor-pair defect $v_{Ga}+Bi_V$, which involves a vacancy and an isovalent substitution of a smaller group V atom with larger Bi atom. A possible reason for this is that the strain induced by the Bi atom is partially relieved by the neighboring vacancy. The effect diminishes as the Fermi level increases due to the introduction of additional electrons, compensating for the missing electrons due to the va-cancy. This compensation is directly connected with the decrease of the formation energy of $v_{Ga}$ point defect in n-type materials. The same effect is not as pronounced in GaSb:Bi, where the strain effects resulting from size mismatch are small relative to GaAs and GaP, as discussed above. The very high attractive binding energy of $v_{Ga}$+$Bi_V$ pairs in GaP:Bi and GaAs:Bi strongly suggests that unwanted Ga vacancies are more likely to be present when GaP and GaAs are grown with Bi atoms to form an alloy, compared to GaSb:Bi which again is predicted to have the lowest tendency to form unwanted defects.

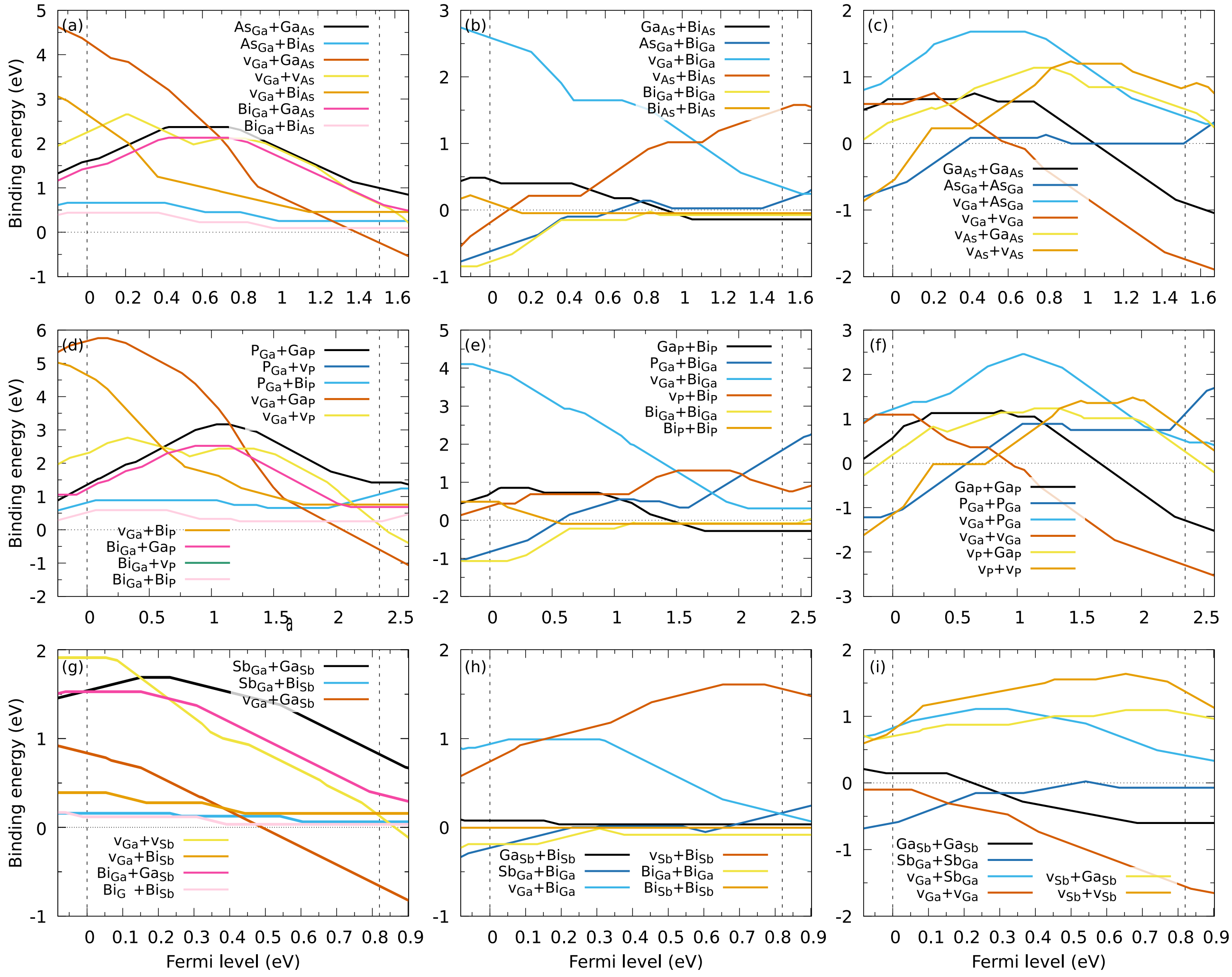

FIG. 2. Binding energies for pair defects in Ga-V:Bi compounds, calculated with the HSE functional. a)-c), d)-f), and g)-i) correspond to GaP:Bi, GaAs:Bi, and GaSb:Bi respectively. Each material has been separated into three columns for clarity, where the first column includes the nearest-neighbor-pair defects, the second contains second-nearest-neighbor-pair defects with Bi related defects, and the third contains the remaining second-nearest-neighbor defects.

In the case of in-situ or ex-situ annealing, which is widely applied to HMAs, it can be expected that the formation/annihilation of point defects will be strongly influenced by formation energies as well as the kinetics of each defect type. Therefore, the possibility of clustering, which is very often suggested to occur in HMAs, including dilute bismides, is worth investigating. In the above discussion, we concluded that the $Bi_V$ atom clustering tendencies in GaSb:Bi differs significantly from that in GaP:Bi and GaAs:Bi since the binding energy for $v_{Ga}$+$Bi_V$ defect pairs in these alloys is much larger (i.e. attractive) than in GaSb:Bi, (Fig. 2). This conclusion is consistent with the experimental data reported so far for these alloys. For example, for GaAs:Bi a strong clustering upon annealing was reported in Refs. [57, 58] while a very homogeneous alloy was observed for GaSb:Bi [59, 60]. One of the pair defects, namely $Bi_{Ga}$+$v_V$ was found to be unstable in all charge states - it undergoes a structure change to a more stable form of $v_{Ga}$+$Bi_V$. A similar situation is observed in the case of negatively charged $V_{Ga}$+$v_V$ in GaP:Bi and GaAs:Bi, where a more stable $v_{Ga}$ is observed.
It is also interesting to note the fact that the binding energy curves as a function of Fermi energy are not always flat. This shows that in many cases the pair defects are not in the same charge states as the point defects comprising the pair. Furthermore, the curves are not always convex because the pair and corresponding point defects change stable charge states as a function of Fermi energy in different ways.

Overall, the calculated database in this work provides significant insights into the behavior of defects in the Ga-V:Bi (as well as In-V:Bi and Al-V:Bi) systems. We expect that as more experimental data on defect levels in dilute bismides becomes available, particularly from DLTS, the present data will provide a valuable resource to aid in interpretation of experimental results, enhance understanding of III-V:Bi defect properties and, subsequently, aid in materials design and optimization. Additional details on results for the remaining In-V:Bi and Al-V:Bi systems obtained with the correction schemes described in Sec. II C can be found in Sec. II C 2.

### C. Modified band alignment method

The band alignment (BA) methods described in Methods section (Sec. IV B) have been used and tested in previous studies on a few select systems, and the mean absolute error between the values obtained with the BA correction and full HSE calculations has been found to be 0.24 eV and less than 0.2 eV in Refs. [40] and [39], respectively. However, in our case, the number of studied defect types and charge states is significantly larger than in previous studies. Here, we have 188 stable defect levels compared to around 20 and 55 in [40] and [39], respectively. This large number of defects allows for a more quantitative assessment of the method. The applicability of the method is apparent from comparing Figs. 3b) and e). In Fig. 3b), which shows the accuracy of LDA vs. HSE for charge state transition levels, a clear underestimation of the defect levels can be observed, with a mean error equal to the negative of the mean absolute error ME = -MAE = $-0.439$ eV. Fig. 3e) shows the LDA vs. HSE charge state transition levels after the BA correction, where the error statistics are much improved with MAE = 0.226 eV and ME = 0.18 eV. See Tab. I for easier comparison of error values. The BA results from Ref. [40] reveal that certain charge state transition levels, in particular those where the charge localization effects are not appropriately described within the local/semilocal functional [61], may show larger inaccuracies and in extreme cases be falsely determined to be unstable. A specific example where these errors might be particularly large is in defects with large Jahn-Teller distortions that require proper charge localization to capture in a quantitative manner. These situations are taken into account and are quantified by precision, recall and F1 scores, which carry information on the amount of misclassified defect levels and formation energies. These values are present on all charge-state transition level and formation energy parity plots.

Although the BA provides a significant improvement, reducing the mean error values by a factor of 2, the now positive value of the mean error indicates the correction overestimated a majority of the values. A visual inspection of Fig. 3e) reveals that the overestimation occurs mostly near the bottom of the band gap, while values at the top of the band gap seem to be more accurate, and that the overestimation is more severe for higher band gap materials. A similar behavior can be observed in Fig. 2 in Ref. [39], where values near the bottom of the band gap tend to be overestimated after the BA correction. These trends suggest an opportunity for improvement of the BA correction scheme. In general, the behavior described above can be remedied by adding two band-gap dependent linear terms to the constant shift:

$$E^d_{corr} = E^d + E_{shift} + \beta\left((1-\delta)E_g - E^d\right). \quad (1)$$

This modified BA method (MBA) results in a shift dependent on the band gap of the material and on the position of a particular defect state within the gap. Such a correction is justified based on the trends apparent in Fig. 3e) and is clearly capable of improving the obtained charge-state transition levels for the data presented here (as can be seen in Fig. 3h)), although its physical interpretation is not obvious. The large number of defects and charge states calculated in this work with both LDA and HSE allows for an empirical determination of the $\beta$ and $\delta$ parameters. In our case, all three Ga-V:Bi material systems have been used in determination of these parameters. Minimizing the RMSE for the general formula (Eq. 1) resulted in $\delta = 0.05$ and $\beta = -0.14$. Given that the value of $\delta$ is so close to zero simply set it to zero, resulting in a simpler one parameter correction formula. With $\delta$ set to zero, the minimization of RMSE resulted in an optimal value of $\beta = -0.14$ and only 1% reduction in RMSE. Further analysis of one- versus two-parameter formulas using Akaike information criterion [62] as well as Bayesian information criterion [63] shows that both criteria support the one-parameter, simplified equation. The MBA method with the optimized value of $\beta$ and $\delta = 0$ leads to a modest but significant improvement in the error statistics of defect levels vs. the BA, producing a MAE = 0.182 eV and ME = 0.001 eV, which is an improvement of about 19% (99%) for MAE (ME) relative to the original band alignment correction method, respectively.

As mentioned in Sec. IVB, the BA correction on the defect levels can be projected to formation energies. As a result, BA with the use of Eq. 6 improves the accuracy of formation energies, reducing the error statistics from MAE = 1.612 eV and ME = $-1.538$ eV for the uncorrected LDA vs. HSE, to MAE = 0.904 eV and ME = $-0.89$ eV. These values are calculated for the formation energies of stable charge states of defects at $E_F = 0$ eV (i.e. the VBM, p-type condition), which are presented in Fig. 3a) and d) for uncorrected LDA vs. HSE and BA correction, respectively. The extent of the improvement provided by the projection (Eq. 6) is, however, limited by the inability of the correction to influence $q = 0$ charge states. This results in a subsequent error in the $q = 0$ charge state defect formation energies (and, therefore, all other charged defect formation energies) in LDA as compared to HSE. This error is inherent to the LDA approach and may be partially due to the LDA inappropriate description of the band positions (and band gap underestimation). As a consequence, improper description of the change in the total energy associated with charge transfer due to charge density reorganization when defects are introduced. The physical mechanism of this behavior is complex and therefore it is not straightforward to fix with a simple physics-based correction.

A similar procedure of projecting the corrected defect levels onto formation energies, although slightly more complex than that in BA, can be carried out for the MBA:

$$E^f_{corr}[X^q] = E^f[X^q] + qE_{shift} + \\ + \beta(E^f[X^0] - E^f[X^q] - q(1-\delta)E_g) + \gamma. \quad (2)$$

The MBA expression for formation energies (Eq. 2) includes new, band gap-dependent, terms. This is a consequence of the new terms in the expression for the defect levels (Eq. 1) compared to BA (Eq. 5). However, despite the MBA providing an improvement over BA defect levels, it suffers from a similar problem as the BA correction in terms of inaccuracy when projected to formation energies. Therefore we propose an additional empirical correction to the MBA ($\gamma$ term in Eq. 2), a constant shift based on the mean error of formation energies, which was determined to be $\gamma = 0.839$ eV. This results in a significant improvement of error statistics of formation energies compared to BA, with MAE = 0.37 eV, a 59% improvement, and ME = 0 eV by construction. As before, these values are calculated for the formation energies of stable charge states of defects at $E_F = 0$ eV, which are presented in Fig. 3g). Parity plots of formation energies for all defects and charge states, including the unstable ones, for all methods can be found in supplementary Fig. S10. It is worth noting, that although the MBA method is based on first principles calculations the newly introduced $\beta$ and $\gamma$ parameters are obtained empirically.

Finally, as a consequence of the improved accuracy in formation energies and charge-state transition levels, errors in the binding energies of pair defects (panels c), f) and i) in Fig. 3) are also reduced. BA improves the LDA vs. HSE MAE = 1.312 eV and ME = −1.240 eV to MAE = 1.064 eV and ME = −1.064 eV. MBA further reduces the MAE to 0.489 eV and ME = −0.099 eV.

It is interesting to note that for the direct LDA and BA the RMSE errors of the binding energies are much lower than expected by simply adding RMSE of formation energies in quadrature ($\sqrt{3}$ times the formation energy RMSE). This results implies that there is significant cancellation in the formation energy errors when these energies are combined in the binding energies, as might be expected. However, this trend does not continue for the MBA. The MBA gives an RMSE = 0.499 for formation energy and RMSE = 0.780 eV for binding energy. The latter is very close to $0.499\sqrt{3} = 0.864$ eV, which is what would be expected by adding the formation energy errors in quadrature. This result demonstrates that after the MBA correction almost no cancellation of errors is obtained in taking the formation energies, supporting that the MBA has effectively used readily available error reduction information. It is worth mentioning that the binding energies, similarly to formation energies, are a function of Fermi energy. Parity plots correspond to $E_F = 0$ eV and $E_F = E_g$, but the same test performed at a different $E_F$ leads to similar error statistics. Finally, some outliers can be seen in the binding energy parity plots 3 c), f) and i). These are a consequence of the LDA (and, therefore, BA and MBA as well) occasionally being unable to predict a certain charge state transition as stable when compared to HSE. This has been discussed in more detail in [40] and in Sec. II C. As a consequence, in some instances, the binding energies of LDA, BA and MBA for one of the point defects or the pair defect are predicted to be in a different charge state than in HSE, leading to a larger error. The residuals (Fig. S11 c)), however, follow a reasonably normal distribution.

In general, comparison of panels a)-c) with d)-f) and g)-i) in Fig. 3 shows that the MBA correction not only brings the results of a pure LDA approach close to the reference HSE functional values, but also improves upon the standard BA method, at the same time preserving the computational efficiency of LDA. This result shows that the HSE defect properties may be predicted reasonably accurately using the results of significantly faster but less accurate LDA methods. This use of two levels of accuracy in the modeling is sometimes called a multifidelity approach and has been previously used with success [46, 64, 65].

#### 1. *Method validation*

The MBA correction (Eqs. 1 and 2) is most useful if its parameters can be readily transferred to a new system. The $\beta$ and $\gamma$ (and $\delta = 0$) parameters in the MBA are fitted to the Ga-V:Bi systems, and not necessarily universal and transferable to other systems. In order to assess their transferability, a new database of single species point defects (i.e. no pair defects) was calculated for the In-V:Bi systems, yielding a set of $3 \cdot 6 \cdot 11 = 198$ defect formation energies. The calculations used the exact same approach as in the Ga-V:Bi calculations and included both LDA and HSE values. The values of $\beta = -0.14$ and $\gamma = 0.839$ eV (and $\delta = 0$) optimized entirely on the Ga-V:Bi dataset were then used for a MBA correction of charge-state transition levels and defect formation energies of point defects in In-V:Bi. The corrected values were compared to the actual HSE results. The magnitude of the correction of our MBA method is scaled by the band gap, so it might be expected to be least effective for low band gap materials. Therefore, performing the validation on the low band gap In-V:Bi was chosen to assess a perhaps worst case scenario applicability of the method. Figs. 4b), e) and h) show defect level parity plots of HSE results with, respectively, LDA, LDA with BA correction, and LDA with MBA correction (Eq. 1, with $\beta = -0.14$ and $\delta = 0$). Similarly as in the case of Ga-V:Bi materials, the BA correction (panel e)) provides a significant improvement over the uncorrected values (panel b)), but exhibits an overestimation in the lower region of the band gap. This, in turn, is remedied by applying the MBA (panel h)), although due to the much lower band gaps of the In-V:Bi systems and therefore fewer stable defect levels, the effect is not as pronounced, and the main improvement is observed through the reduction in mean error. The final error values for the MBA are MAE = 0.175 eV and ME = $-0.009$ eV, which represent 50% and 97% improvement over the uncorrected values of MAE = ME = $-0.35$ eV. This result demonstrates that the empirically obtained $\beta = -0.14$ and $\delta = 0$ are transferable to this system. Figs. 4a), d) and g) show formation energy of defects in stable charge-states at $E_F = 0$ eV parity plots of HSE results with, respectively, LDA, LDA with BA correction, and LDA with MBA correction (Eq. 2, with $\beta = -0.14$ and $\gamma = 0.839$ eV, (and $\delta = 0$)). Their analysis, again, reveals improvement of the error statistics for BA as compared to uncorrected values, and further improvement when MBA is used, confirming the transferability of $\gamma$ and $\beta$. The final error values for the MBA are MAE = 0.384 eV and ME = 0.236 eV, which represent 62% and 74% improvement over the uncorrected values of MAE = 1.014 eV and ME = 0.905 eV. Parity plots of formation energies for all defects and charge states, including the unstable ones, can be found in supplementary Fig. S10.

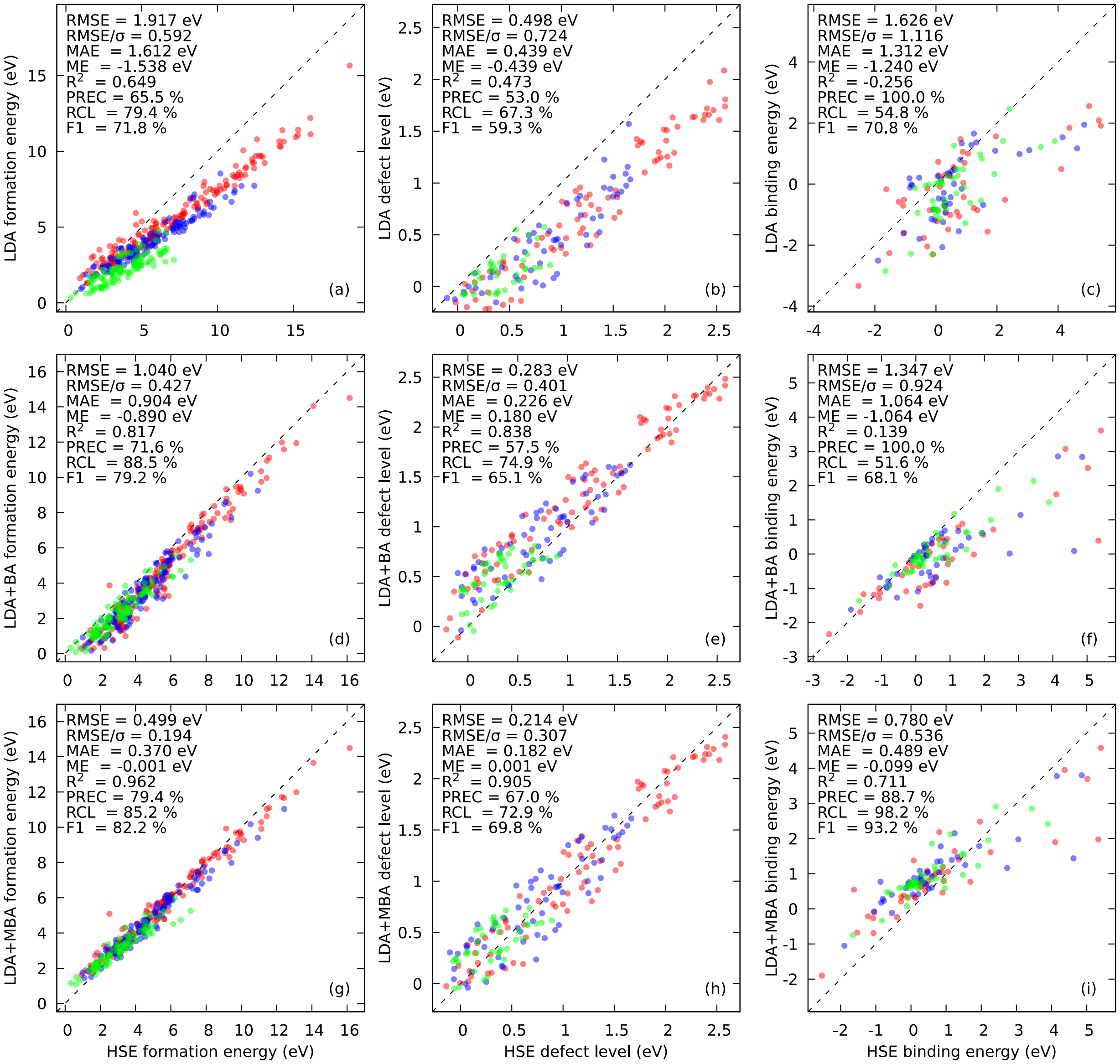

FIG. 3. Accuracy comparison of different approaches with hybrid functional (HSE) results on Ga-V:Bi dataset. First row: baseline LDA vs. HSE, second row: band alignment (BA) correction, third row: modified band alignment (MBA) correction. First column: formation energies of stable defects (precision and recall pertain to whether a defect in a certain charge state is predicted to be stable), second column: charge-state transition levels (precision and recall pertain to whether a charge-transition is properly predicted inside the band gap), third column: binding energies (precision and recall pertain to whether a binding energy is properly predicted as positive or negative). Red, blue and green points correspond to defects in GaP:Bi, GaAs:Bi, and GaSb:Bi respectively.

The In-V:Bi systems are likely to be quite similar with the Ga-V:Bi systems used to obtain the $\beta$ and $\gamma$ values and therefore significant additional study in other materials families, e.g. II-VI systems, is needed to establish the general applicability of the MBA vs. BA method. In order to further demonstrate the applicability of MBA, we apply the method to a set of 656 PBE-calculated charge state transition levels of defects and impurities in CdX (X=S, Se, Te) reported by Mannodi-Kanakkithodi *et al.* [46], and compare them to the equivalent HSE values reported therein. As mentioned in Sec. IV, the BA and MBA methods are, in principle, applicable to semilocal functionals as well, therefore this test will assess not only the transferability of the method and its optimized parameters to other systems, but to other functionals as well. The resulting parity plots can be found in Figs. 4 c), f) and i). The MBA method (panel i)) again proved to be very efficient with MBA predicted values giving a MAE = 0.194 eV and ME = −0.006 eV vs HSE values. This represents an improvement of 16% and 96% compared to the MAE = 0.230 eV and ME = 0.137 eV obtained with pure BA and an improvement of 55% and 99% over the uncorrected defect levels with MAE = 0.427 eV and ME = −0.415 eV. The work in Ref. [46] allows us to make a direct and independent comparison between the MBA and machine learning (ML) methods. Ref. [46] provides a ML model for predicting defect level values as close as possible to HSE from input features that include elemental properties and PBE defect levels. We can therefore compare the errors from the MBA and the ML model where both have the full PBE defect information available. We note that all the data sets in Ref. [46] include charge states outside the gap, which were not used in the optimization of our present MBA model. Ref. [46] obtained a RMSE = 0.24 eV with their best average ML model (Random Forest Regression - RFR) on a 10% left out test set from their main training data. To compare to this, we use the MBA to predict their main data set (including the test set) split into 10 folds, to obtain an average and standard deviation RMSE = 0.240±0.035 eV (Fig. S12 c)). We also compare to their predictions on an out-of-sample test comprised of data on two new systems not directly in their training space ($CdTe_{0.5}Se_{0.5}$ and $CdSe_{0.5}S_{0.5}$). On this data, the ML model in Ref. [46] obtained a RMSE = 0.235 eV and the MBA obtained a RMSE = 0.267 eV (Fig. S12 f)). The MBA RMSEs for the out of sample test set are around 13% worse than the ML model, which is statistically significant, but the MBA provides a much simpler approach, with just one fitting parameter for defect levels that appears to be quite transferable. Additionally, it is important to notice that even though the MBA was optimized on LDA calculated values, the optimized parameters also apply to the PBE, and it is expected that they should apply to other semilocal functionals as well. All of the error statistic values are gathered in Tab. I.

The CdX (X=S,Se,Te) results from Mannodi-Kanakkithodi *et al.* [46] were the only large readily available database for defect levels calculated with both hybrid and local/semi-local functional, which is necessary to critically assess the transferability of the MBA method. Furthermore, this database shared many features with ours, including the same crystal structure and similar methods of calculations (e.g. the same supercell size, DFT code and electrostatic correction), providing perhaps near optimal condition for transferability of our MBA. We have performed some more limited but more demanding testing for transferability with the much smaller set of data reported in [39] and [40]. The results are shown in supplementary Fig. S14. The MBA for dataset from Ref. [39] shows an improvement of 20% in MAE over the regular BA, however the parameters need to be reoptimized, and both $\beta$ and $\delta$ need to be utilized. The MBA method used on, another smaller dataset, reported in [40], results in an improvement in MAE of 8% over the regular BA with the parameters reoptimized. Both of these smaller datasets are, however, performed on larger supercells, in systems of different crystal structure and chemistry and with slightly different computational methods, which is potentially the reason why the reoptimization is necessary. Although the need to reoptimize parameters may make the method more time consuming to apply in some cases, analysis performed on our main dataset (Ga-V:Bi systems) shows that using as few as 10 points for reoptimization produces results that already show noticeable improvement over the regular band alignment method. Supplementary Fig. S15 shows the MAE as a function of the number of points used in parameter optimization. The rapid convergence of the fitting means that the MBA can be refit with very modest amounts of data. These validation tests show that the modified band alignment method does improve the results of the regular band alignment approach in all cases studied here, and those cover a wide variety of different systems of different chemistry and different computational approaches, which suggests that the method is likely to be widely applicable. However, the parameters may need to be reoptimized for different unit cell sizes and crystal structures than those that were used in this study or in Ref. [46]. For calculations similar to those performed here or in the CdS study [46] our MBA correction with $\beta = -0.14$ and $\gamma = 0.839$ eV (and $\delta = 0$) may be applied directly to modify the regular band alignment method and provide a quick path to increasing the fidelity of LDA or GGA defect levels and formation energies to approach HSE accuracy.

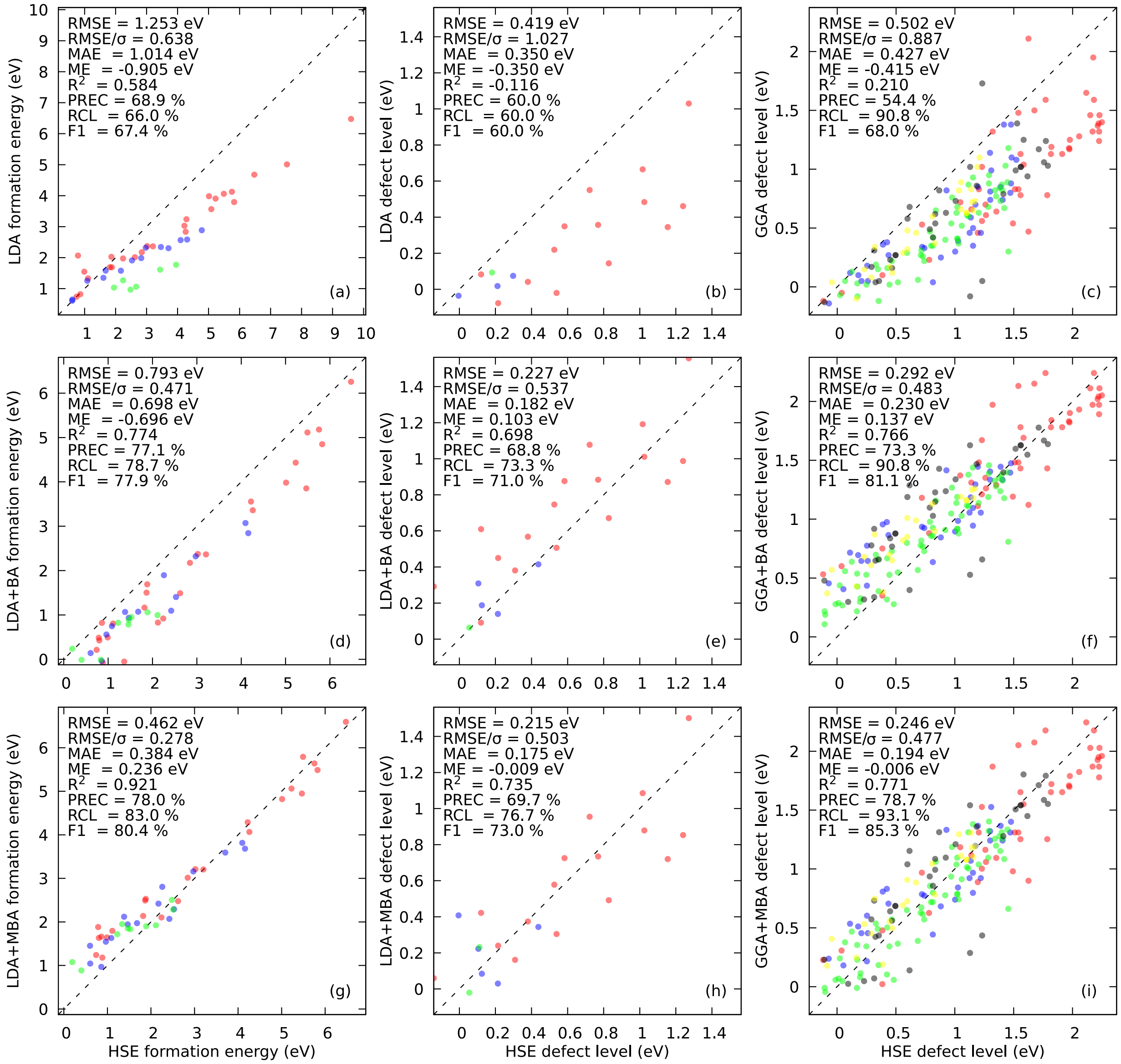

FIG. 4. Validation of the correction methods for In-V:Bi point defects and a test set of CdX (X=S, Se, Te) defects and impurities [46], optimized on Ga-V:Bi dataset. First row: baseline LDA/GGA vs. HSE, second row: band alignment (BA) correction, third row: modified band alignment (MBA) correction. First column: formation energies of stable defects (precision and recall pertain to whether a defect in a certain charge state is predicted to be stable), second column: charge-state transition levels (precision and recall pertain to whether a charge-transition is properly predicted to be stable and inside the band gap). Red, blue and green points correspond to defects in InP:Bi, InAs:Bi, and InSb:Bi (CdS, CdSe, CdTe) respectively. Grey and yellow are CdSSe and CdSeTe alloys.

### 2. *Application of the method to new systems*

After demonstrating that the MBA is an effective correction of defect properties (see Sec. II C 1), the MBA was subsequently used to obtain defect properties of the remaining (pair) defects in the In-V:Bi family of materials as well as all 27 defect types in Al-V:Bi. These predictions, combined with the directly calculated HSE results for Ga-V:Bi and point defects in In-V:Bi resulted in a large database of 2673 defect formation energies. These systems are all naturally zincblende III-V semiconductors which are technologically relevant for optoelectronic applications. The calculated values included the formation energies, binding energies, and defect charge-state transition levels. Due to the mixed direct HSE DFT and LDA+MBA correction approaches in tables and figures, MBA results are marked with an asterisk. Tables II and III include the MBA defect levels for In-V:Bi materials alongside the directly HSE-calculated Ga-V:Bi, while the MBA values for Al-V:Bi can be found in Sec. VI A, Tab. SII). The MBA binding energies for In-V:Bi and Al-V:Bi are present in Figs. 6 and 5 respectively. Point defect formation energies are collected in one figure together with the directly calculated values in Fig. 1. Due to the large amount of calculated data, only point defect formation energies in intermediate conditions together with binding energies are presented here. The formation energies of pair defects can be estimated from the binding energies, or can be found in the supplementary material (Sec. VI A, Figs. S1-S9) for both group III and group V rich conditions, as well as intermediate ones.

TABLE I. Comparison of error statistics of all the studied methods on different test data.

| | | Formation energy | | | Defect Levels | | | Binding energy | | |
|---|---|---|---|---|---|---|---|---|---|---|
| | | Baseline (LDA vs. HSE) | BA | MBA | Baseline (LDA vs. HSE) | BA | MBA | Baseline (LDA vs. HSE) | BA | MBA |
| **Ga-V:Bi** | RMSE (eV) | 1.917 | 1.040 | 0.499 | 0.498 | 0.283 | 0.214 | 1.626 | 1.347 | 0.780 |
| | MAE (eV) | 1.612 | 0.904 | 0.370 | 0.439 | 0.226 | 0.182 | 1.312 | 1.064 | 0.489 |
| | ME (eV) | -1.538 | -0.890 | -0.001 | -0.439 | 0.180 | 0.001 | -1.240 | -1.064 | -0.099 |
| **In-V:Bi** | RMSE (eV) | 1.253 | 0.793 | 0.462 | 0.419 | 0.227 | 0.215 | - | - | - |
| **test set** | MAE (eV) | 1.014 | 0.698 | 0.384 | 0.350 | 0.182 | 0.175 | - | - | - |
| | ME (eV) | 0.905 | -0.696 | 0.236 | -0.350 | 0.103 | -0.009 | - | - | - |
| **CdX [46]** | RMSE (eV) | - | - | - | 0.502 | 0.292 | 0.246 | - | - | - |
| **test set** | MAE (eV) | - | - | - | 0.427 | 0.230 | 0.194 | - | - | - |
| | ME (eV) | - | - | - | -0.415 | 0.137 | -0.006 | - | - | - |

The chemical trends observed in the results of the predicted values for In-V:Bi and Al-V:Bi materials are generally analogous to the trends observed for the directly calculated values for Ga-V:Bi, discussed in Sec. II. Point defect formation energies exhibit similar trends with mismatch, with the most encouraging properties visible for III-Sb:Bi. The differences in group III and Bi atoms size, although less prominent, are also visible in the chemical trends but do not influence the general conclusions, apart from the fact that $Bi_V$ charge-transition transition levels are visible above the VBM for all Al-V:Bi materials. The main differences visible in the shape of the formation energies and binding energies come from the large difference in the band gaps between Al and In based materials.
The results of all calculations can be found in VI B.

### D. Machine learning model exploration

Recent studies on machine learning (ML) for impurities in Cd based chalcogenides [46] showed that ML can be a powerful tool for efficient predictions of defect properties based not only on the results of less expensive semi-local functionals, but even just elemental properties of the impurities, greatly reducing the effective computational cost. The notable success of the relatively simple BA and MBA methods suggests that a simple relationship might exist between the LDA and HSE defect energetics which could be effectively captured with machine learning. To explore this hypothesis, we consider the regression problem of fitting $F$ in $Y = F(X)$, where the target $Y$ are the HSE defect formation energies and features $X$ are data we can obtain from LDA. Note that the $X$ data could be just defect formation energies, but we could include some other features that come at no extra computational cost from the LDA calculations. Tools used in the machine learning are described in Sec. IV D. The $X$ features used in the regression included the formation energy, charge state, band gap, FNV charge interaction correction, total energy, and results of Bader analysis (which captures aspects of localization effects). We first considered linear multivariable regression, which revealed the charge state and the formation energy to have by far the highest coefficients. This result suggested that the band alignment shift is the major factor in the discrepancy between HSE and LDA results. Consistent with this result, the validation on In-V:Bi test data, equivalent to that in Fig. 4, resulted in statistics very similar to those of the MBA method. Second, in order to try more complex nonlinear methods, other machine learning approaches were optimized to yield the lowest leave-system-out cross validation (CV) MAE. Neural networks yielded the most promising results. After a modest effort of testing the number of layers and nodes, a neural network of 3 layers with 128, 64, and 32 nodes was found to be fairly optimal. As a result of leave-system-out CV, the obtained error statistics were ME = 0.04 eV and MAE = 0.195 eV for the defect levels and ME = $-0.115$ eV and MAE = 0.343 eV for the formation energies. Random 5-fold CV yielded only marginally better error statistics. Supplementary figure Fig. S13 shows parity plots of the results of the leave-system-out CV of the model vs. HSE results of formation energies and defect levels. The result is not noticeably better than the MBA method (Fig. 3 g), h)), but required a much more computationally intensive machine learning algorithm and training on both HSE and LDA datasets, making the method much less accessible and more difficult to apply. Additionally, due to simplicity and being more physics-based, MBA is expected to be much more transferable to new systems.

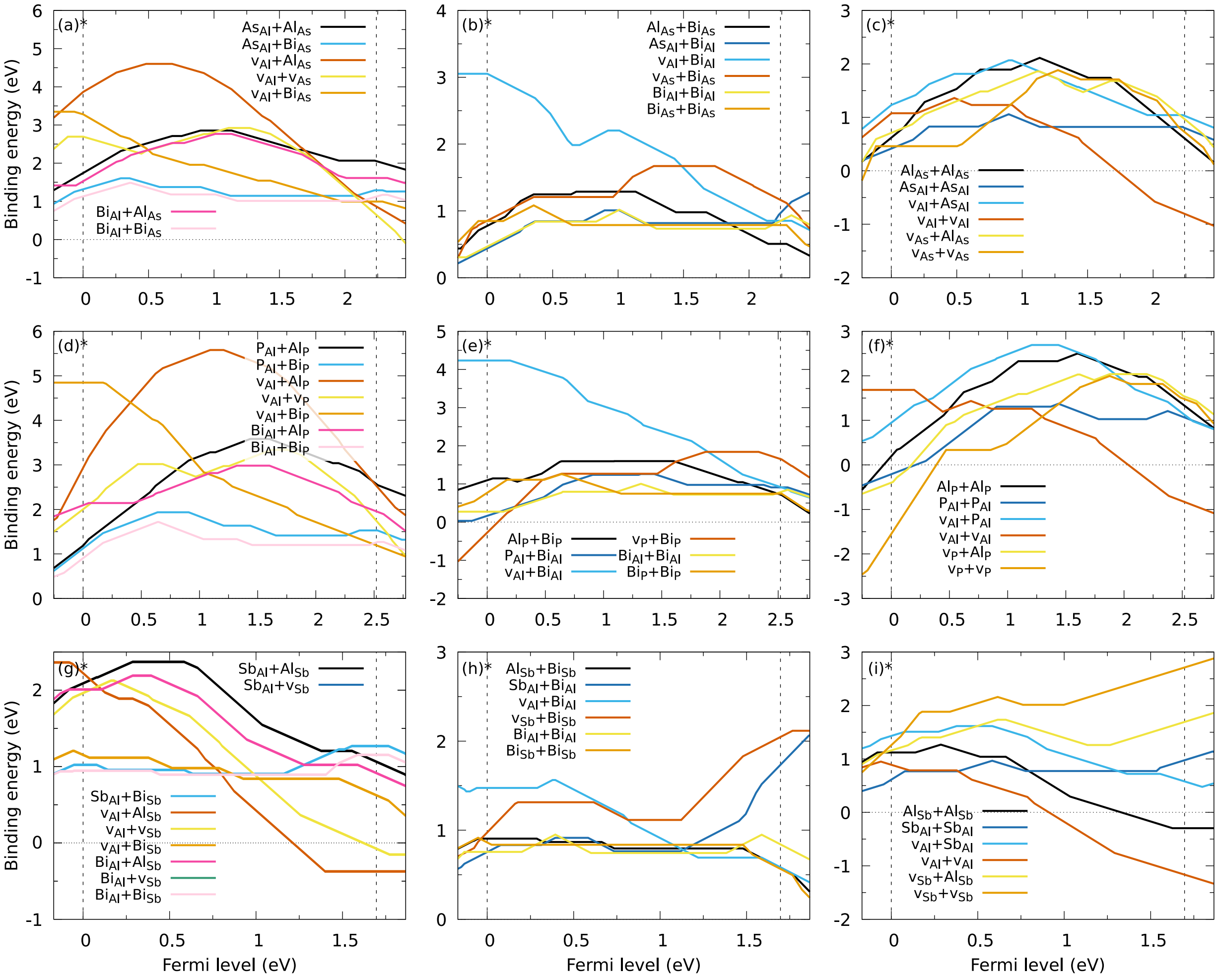

FIG. 5. Binding energies obtained with the modified band alignment method (MBA) for pair defects in Al-V:Bi compounds. a)-c), d)-f), and g)-i) correspond to AlP:Bi, AlAs:Bi, and AlSb:Bi respectively. The asterisk indicates that the results were obtained with LDA corrected with the modified band alignment correction (MBA).

Therefore, based on the results obtained here as well as the trends of local/semilocal defect levels vs. HSE observed in Refs. [39, 40, 46], we believe that the MBA method may be the most practical approach.

It is worth to notice that in case of impurities, as demonstrated in [46, 66], ML methods may be able to provide approximate description of defect levels and formation energies solely from elemental properties. However, for the present study which focuses mostly on intrinsic defects, such type of ML model was not possible to explore.

## III. SUMMARY

In this work, we provide a comprehensive, self-consistent database of defect formation energies, charge-state transition levels and binding energies for point and pair defects using state-of-the-art *ab initio* methods. The computed database covers all zincblende III-V diluted bismides, which are technologically relevant for a number of optoelectronic applications. Based on the obtained results, band alignment and machine learning approaches are investigated to obtain hybrid functional accuracy defect formation energies and defect levels from more inexpensive functionals. A new method of correcting results of computationally inexpensive functionals, the modified band alignment, is proposed and assessed in detail. This research provides valuable information for a range of materials research modalities. The large database can aid experimental researchers in identification of defects observed in experimental measurements and provide rational strategies to tune the defect properties of a material. The large amount of data can be directly used by the materials informatics community to design, build and assess new machine learning models for more quantitative prediction of defect properties and understanding of chemical trends in a range of semiconductor systems. Finally, the modified band alignment method proposed here is directly useful for computational researchers conducting atomistic simulations of de-fect properties. The method enables prediction of defect formation energies and charge state transition levels with accuracy approaching that of hybrid HSE functionals but at the computational expense of LDA/GGA calculations. Our proposed method has been thoroughly verified on our own tests, as well as on defect data for an entirely different family of materials obtained from a separate study. The modified band alignment method not only allows for fast evaluation of defect properties, but also opens up new opportunities for high-throughput calculations of defect and related properties in other systems.

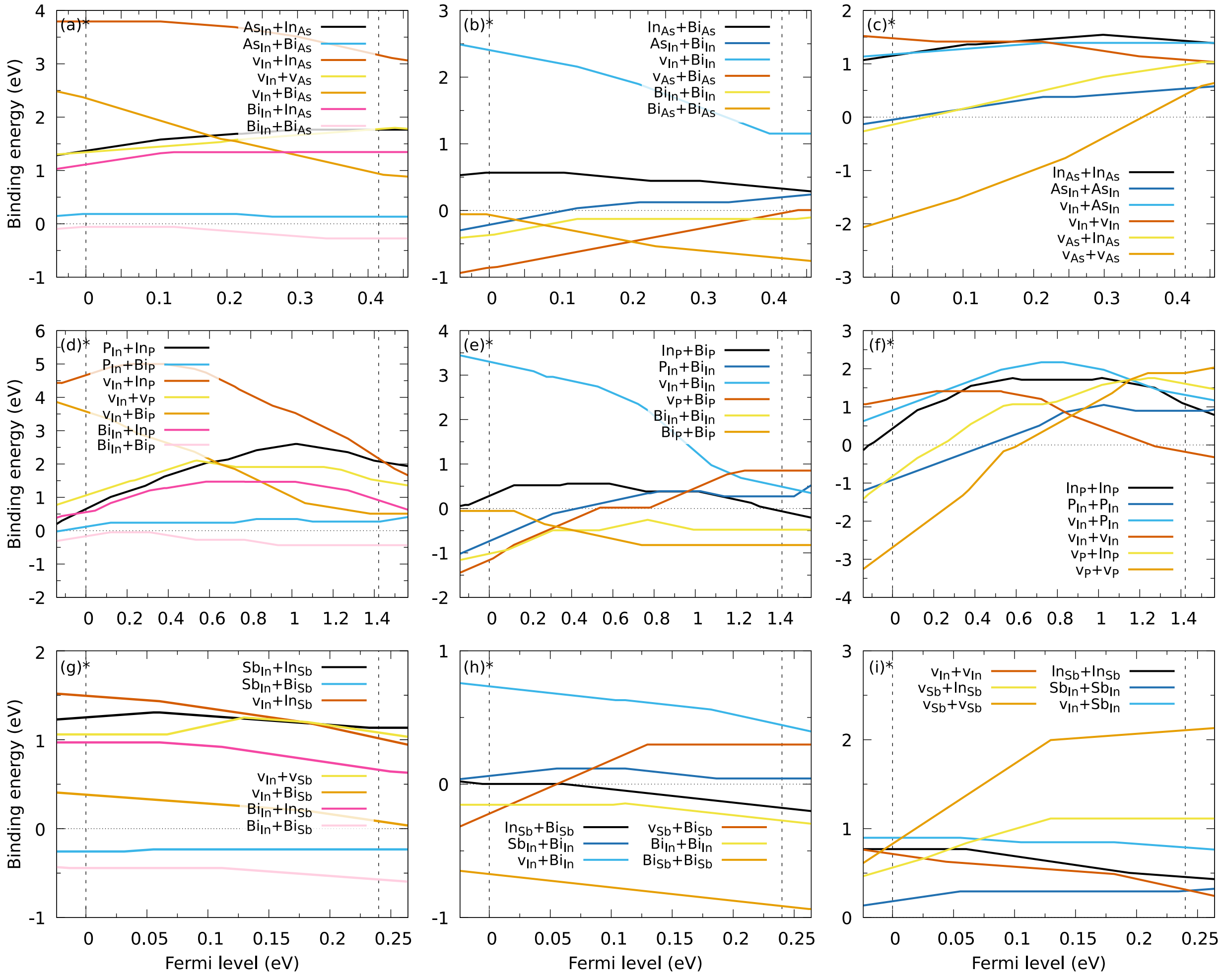

FIG. 6. Binding energies obtained with the modified band alignment method (MBA) for pair defects in In-V:Bi compounds. a)-c), d)-f), and g)-i) correspond to InP:Bi, InAs:Bi, and InSb:Bi respectively. The asterisk indicates that the results were obtained with LDA corrected with the modified band alignment correction (MBA).

## IV. METHODS

### A. Defect properties

For each material, 27 different types of defects were studied. 6 point defects: $Bi_{III}$, $Bi_{V}$, $v_{III}$, $v_{V}$, $III_{V}$, $V_{III}$, and 21 pair defects (the defects are described in Kroger-Vink notation where III and V refer to elements from these columns of periodic table, and v refers to vacancies). The pair defects included all possibilities of nearest- and second-nearest-neighbor pairs of point defects. Charges ranging from -5 to 5 were analyzed. As no defects of charge -5 (5) were found to be stable, no higher charged defects were investigated.

The formation energy of a defect $X$ with a charge $q$ added to the computational unit cell as a function of Fermi level $E_F$ in the band gap was calculated according to the formula:

$$E^f[X^q](E_F) = E_{tot}[X^q] - E_{tot}[\text{pure}] - \sum_i n_i\mu_i + \\ + q(E_{VBM}[\text{pure}] + E_F) + E_{corr} \quad (3)$$

where $E_{tot}[X^q]$ is the total energy of the supercell with the defect, $E_{tot}[\text{pure}]$ is the total energy of the corresponding pure (undefected) supercell, $n_i$ is the number of added ($n_i > 0$) or removed ($n_i < 0$) atoms, $\mu_i$ is the chemical potential of the species $i$, $E_{VBM}[\text{pure}]$ is the energy of the valence band maximum of the pure material and $E_{corr}$ is the energy correction that compensates for the electrostatic interactions arising from the periodic image of the defect. The Freysoldt, Neugebauer and Van de Walle (FNV) correction [67, 68] was used to account for the electrostatic interaction.

For pair defects $XY^q$ comprised of point defects $X_q$ and $Y^q$, the binding energy was evaluated according to:

$$E_b(E_F) = E^f[X^{q_X}](E_F) + E^f[Y^{q_Y}](E_F) - E^f[XY^{q_{XY}}](E_F) \quad (4)$$

where $q_X$, $q_Y$ and $q_{XY}$ correspond to the charge state of the particular defect with the lowest formation energy at the chosen Fermi level. Positive values of binding energy indicate an energetic preference of the defects forming as a pair rather than separately.

A stable charge state is define by the lowest formation energy $E^f[X^q](E_F)$ at a given Fermi level $E_F$. The energies at which the resultant curve changes slope (intersections of $E^f[X^q](E_F)$ lines) are the charge state transition (defect) levels, $E_d$.

It is worthwhile to mention here that the $Bi_V$ is an isovalent dopant, which is usually not treated as a defect. However, since it may influence the electronic structure and have a similar physical effect on the system as other defects, treating it as such in this work provides a much more convenient way of comparing and presenting the results.

### B. Band alignment based correction of LDA results

One of the most efficient and at the same time computationally inexpensive methods of correcting the LDA results of charge-state transition levels (and, consequently, formation energies) is the band alignment correction family of methods [39–45]. In these methods, a common reference for the charge-state transition levels is established and allows one to align the band edges between the semilocal/local and hybrid functional calculations. Over the years, a number of different approaches have been used: Ref. [39] used local ionic potential as a reference, Ref. [41] used a deep $2s$ atomic level, Ref. [40] utilized slab calculations to allow the use of vacuum as a reference, and Refs. [42–45] used the average electrostatic potential. In this work, we adopt the last approach, where the alignment is obtained from the average electrostatic potential. The main advantage of this method is its simplicity and the requirement of only one HSE calculation on the primitive unit cell of the undefected system, effectively making the correction quick and easy to apply. Although the band gaps calculated within LDA or GGA and HSE differ significantly, the geometry of the primitive unit cell calculated within LDA or GGA and HSE are often similar, which is also true in the case of materials studied in this work. Therefore, we assume that the structural differences have negligibly small impact on the electronic densities which is necessary for the method to succeed [42]. This allows for the method to be used on fully self-consistent LDA or GGA calculations and validated against independent fully self-consistent HSE results. In practice, in this work, the band alignment is obtained by first calculating the electronic structure and electrostatic potential of the pure material with both LDA and HSE, and then calculating the difference between the averaged electrostatic potentials (band alignment shift, $E_{shift}$). The value is then added to the LDA charge state transition levels (the band edges are *aligned* according to the average electrostatic potentials). This results in a constant shift of those levels on the energy scale.

$$E^d_{corr} = E^d + E_{shift} \quad (5)$$

where $E^d$ is the LDA defect level. It is worth noting that the underestimated value of the LDA band gap is also corrected in this method and the systems are analyzed and interpreted within the correct, and appropriately aligned, HSE band gap. After the alignment the defect levels are referenced to the VBM of the corrected band gap. A shift of a charge-state transition level, i.e. the intersection of two $E^f(E_F)$ (Eq. 3) lines associated with two charge states, may be interpreted as shifts of the $E^f(E_F = 0)$ formation energy levels of those two charge states. Therefore, correction associated with the band alignment shift of the defect levels can be projected onto the formation energies, resulting in a charge-dependent correction of the $E^f(E_F = 0)$ formation energies.

$$E^f_{corr}[X^q] = E^f[X^q] + qE_{shift} \quad (6)$$

There is, however, a fundamental limitation of this approach to obtaining corrected formation energies. The projection of band alignment correction on the formation energies is charge dependent and, therefore, provides no correction for charge $q = 0$. Although the band alignment projection does significantly improve the accuracy of the formation energies as can be seen by comparing Figs. 3 a) and d), there is still a clear but consistent underestimation, associated with the inability of the correction to affect $q = 0$ formation energies.

### C. First principles calculations

Calculations were performed using density-functional theory [69] as implemented in the `VASP` code [70, 71], and with plane-augmented wave (PAW) potentials [72] with $s2p1$ and $s2p3$ valence electron configuration for group III and V atoms respectively, with the exception of Bi atoms where, due to its large size, $d$ electrons were additionally included. A 64 atom supercell was used, which is a $2\times2\times2$ multiplication of a conventional zincblende unit cell. Convergence studies on GaAsBi [47] have shown that for a $-3$ charge state this choice of supercell, when used in conjunction with the FNV electrostatic correction scheme, leads to an error in formation energies of up to 0.16 eV when compared to a large, 512 atom unit cell. For defect levels, additional tests we performed here led to an estimated error of using 64 atom unit cells not exceeding 0.16 eV (See Fig. S16, where a convergence of defect levels with respect to unit cell size for $v_{As}$ in GaAs for charge states ranging from 0 to $-4$ is presented). It is important to notice that very few defects for the systems studied here are stable in charge states higher than $+-3$ (less than 5%), and the majority of stable states are in the $\pm 2$ range. Considering the computational cost required for the large number of calculations and the HSE hybrid functional [73], which has been used for both the geometry optimization and the total energy calculations, the 64 atom unit cell has been chosen as a compromise between computational cost and accuracy. In addition, the use of 64 atom unit cell allowed for consistency in the additional verification on CdX systems from Ref. [46], which were obtained on 64 atom supercells as well. The internal atomic degrees of freedom were optimized for each defect and each charge state to allow the possibility of different relaxations, including Jahn-Teller distortions. The use of the HSE functional was deemed necessary for a number of reasons: first, to reproduce the intricate electron density of charged structures as accurately as possible, second, to get an accurate value of the total energy, and finally to provide a correct description of the band structure and the band gap value in particular. The $\alpha$ in the HSE was used as a free parameter to fine-tune the obtained value of the band gap for pure parent GaP, GaAs and GaSb compounds, where slight adjustments provided an excellent agreement of band gaps and lattice parameters with those of 0K experimental values. The values of the $\alpha$ parameter used for each material together with the resulting band gaps, lattice parameters and $E_{shift}$ values used in Eqs. 1, 2, 5 and 6 have been summarized in supplementary Table SI. All defect calculations were performed with the optimized $\alpha$ values for each material. Convergence studies and assessment of the required computational resources resulted in a choice of $2 \times 2 \times 2$ Monkhorst-Pack mesh [74]. An energy cutoff of 350 eV (1.35 times the recommended value in the POTCAR of the hardest atom) was used, and the total energy within each SCF cycle was converged to 0.1 meV. The optimization procedure was carried out until none of the forces exceeded 0.005 eV/Å. Due to the large mass of Bi and in order to properly reproduce the band structure (the band gap in particular), all calculations were performed with spin-orbit coupling included. It has been shown that the inclusion of spin-orbit coupling has significant influence on the electronic structure as well as on formation energies [47] in GaAs:Bi and the same is expected for other systems studied here. In all cases, the magnetic moments were optimized from initial values of $m_x = m_y = m_z = 1$ Bohr magneton per atom with the assumption that the most stable spin state would be found through optimization. Due to the large scale of the study it was not feasible to explicitly study all possible spin states for all defects, however, since the typical spin polarization energies for similar systems (specifically, vacancies in other similar III-V systems), of present at all, are in the order of tens of meV [75], the error would be negligible.

The shortcoming of the periodic approach to charged defect calculations, specifically, the problem of the electrostatic interaction between the artificially high concentration of defects created by the periodic images of the defects [76], was overcome by the use of the Freydsoldt (FNV) method [67, 68]. The correction was calculated via the alignment of the local potentials of defected and pure structures with the use of the `sxdefectalign` code, an add-on to the `SPXInX` repository. To keep the results consistent throughout the large number of cases studied, the alignment region of the potential was kept consistent throughout all the structures and types of defects. The region was chosen as 30% of the area of the potential in the middle between the periodic images of the defect, from which an average value was calculated and used in the alignment procedure. Dielectric constants are required for proper description of screening properties in the determination of the electrostatic correction. Out of convenience and due to their wide availability for a large number of semiconductor materials, experimental values of dielectric constants were used. Those values were taken from Ref. [77]. For unknown systems, however, our tests have shown an LDA-calculated dielectric constants should yield satisfactory results of the electrostatic correction, with the difference between the energy correction calculated with experimental and LDA dielectric constant not exceeding 10% for a +/-1 charge. The chemical potentials have been calculated for each element as a total energy per atom in their corresponding lowest energy crystal structures, i.e. $Cmca$ for Ga, $I4/mmm$ for In, and $R\bar{3}m$ for all group V elements except phosphorus, which according to both our calculations and Ref. [78] has an equilibrium structure of $P2/c$. The calculation parameters for chemical potentials were kept consistent with the parameters for calculation of formation energies. In the case of HSE, $\alpha$ values corresponding to that used for the host material were used for calculation of chemical potentials. This was done in order to stay consistent and take advantage of potential error cancellation. Although hybrid functionals are not necessarily the most accurate for calculating cohesive energy of metals, they still tend to perform well [73]. The obtained chemical potentials were used in the calculation of defect formation energies, which have been evaluated in III-rich and V-rich conditions (upper and lower bounds). For III-rich conditions, $\mu V$ was obtained with the use of the DFT values of $\mu III$ and $\mu(III\text{-}V)$ and for V-rich, $\mu III$ was calculated respectively with the DFT values of $\mu V$ and $\mu(III\text{-}V)$, in both cases utilizing the relation $\mu III + \mu V = \mu(III\text{-}V)$. Intermediate conditions correspond to chemical potentials calculated as an average between III- and V-rich values. Values of all used chemical potentials can be found in

VI B.
The post-processing of the results of formation energies to obtain binding energies and defect levels from the DFT results and band alignment and modified band alignment approaches was performed with self-written `python` codes. See Sec. VI B for information about the obtained data.

### D. Machine learning models

The machine learning models were built and validated with the MAterials Simulation Toolkit for Machine Learning `MAST-ML` utility [79], which uses numerical procedures as implemented in `scikit-learn` [80].

### E. Appendix: Definitions of statistical quantities

$$\mathrm{ME} = \frac{\sum_{i=1}^{n} y_i - x_i}{n}$$

$$\mathrm{MAE} = \frac{\sum_{i=1}^{n} |y_i - x_i|}{n}$$

$$\mathrm{RMSE} = \sqrt{\frac{\sum_{i=1}^{n} (y_i - x_i)^2}{n}}$$

$$\sigma = \sqrt{\frac{1}{n}\sum_{i=1}^{n} (x_i - \bar{x})^2}$$

$$\bar{x} = \frac{1}{n}\sum_{i=1}^{n} x_i$$

$$\mathrm{Prec} = \frac{true\ positives}{true\ positives + true\ negatives}$$

$$\mathrm{Rcl} = \frac{true\ positives}{true\ positives + false\ negatives}$$

$$\mathrm{F1} = \frac{2 \cdot \mathrm{Prec} \cdot \mathrm{Rcl}}{\mathrm{Prec} + \mathrm{Rcl}}$$

$$\mathrm{R}^2 = 1 - \frac{\sum_{i=1}^{n} (y_i - x_i)^2}{\sum_{i=1}^{n} (x_i - \bar{x})^2}$$

where $y_i$ are the corrected values and $x_i$ are the reference HSE results.

## ACKNOWLEDGMENTS

M. P. P. acknowledges support within ETIUDA grant no. 2016/20/T/ST3/00258 from the National Science Centre (NCN) Poland. The author R. K. acknowledges financial support from NCN (grant no. 2012/07/E/ST3/01742). The authors D.M. and R.J. acknowledges support from the National Science Foundation (NSF) Cyberinfrastructure for Sustained Scientific Innovation (CSSI) (OAC-1931298) . I.S. acknowledges support from NSF through the University of Wisconsin Materials Research Science and Engineering Center (DMR-1720415). Calculations have been carried out using resources provided by Wroclaw Centre for Networking and Supercomputing (http://wcss.pl).

## V. CONTRIBUTIONS

M. P. P. performed the calculations and prepared/analysed the results, D. M. guided and supervised the research. Interpretation of results and writing of the manuscript was done by M. P. P. and D. M.. R. K., R. J. and I. S. revised the manuscripts and discussed the results. All authors read, revised and approved the final manuscript.

## VI. ADDITIONAL INFORMATION

### A. Supplementary Material

Due to the large number of calculations, many figures and tables would not fit into the paper and are all gathered in a supplementary material document published along with this paper on the publisher website.

### B. Data Availability

All datasheets containing results supporting the findings of this study (formation energies, change state transition levels) are available on figshare [81]: https://doi.org/10.6084/m9.figshare.12478700. Raw data (input and output files), excluding POTCARs due to VASP license restrictions, are also included together with the code that allows to postprocess it in order to obtain the results presented in the paper.

### C. Code Availability

The DFT calculations have been performed using VASP (v5.4.4) [70, 71], the electrostatic correction has been calculated using the `sxdefectalign` (v2.2) [67] which is available directly from the authors at: https://sxrepo.mpie.de/projects/sphinx-

TABLE II. Charge-state transition levels in reference to the valence band maximum. The levels printed in gray lay outside of the gap but fit within 10% of the value of the gap. Numbers in parenthesis indicate the charge-state transition. Energy values are given in electronvolts (eV). Columns calculated with the MBA approach are indicated with an asterisk.

| | Point defects | | | | | | Nearest neighbour pair defects | | | | | |
|---|---|---|---|---|---|---|---|---|---|---|---|---|
| Defect | Bi$_{Ga}$ | Bi$_P$ | Ga$_P$ | v$_{Ga}$ | P$_{Ga}$ | v$_P$ | Bi$_{Ga}$+Bi$_P$ | Bi$_{Ga}$+Ga$_P$ | v$_{Ga}$+Bi$_P$ | v$_{Ga}$+v$_P$ | P$_{Ga}$+Bi$_P$ | P$_{Ga}$+Ga$_P$ |
| GaPBi (eV) | 0.64 (2/1)<br>1.16 (1/0) | 0.07 (1/0) | 0.09 (2/1)<br>0.31 (1/0)<br>0.87 (0/-1)<br>1.14 (-1/-2) | -0.14 (1/0)<br>0.16 (0/-1)<br>0.76 (-1/-2)<br>1.06 (-2/-3) | 1.04 (2/1)<br>1.36 (1/0)<br>2.52 (0/-1) | 0.33 (3/1)<br>1.34 (1/0)<br>1.51 (0/-1)<br>1.93 (-1/-2)<br>2.02 (-2/-3) | 0.90 (2/1)<br>1.23 (1/0)<br>2.37 (0/-1) | -0.08 (4/2)<br>0.17 (2/1)<br>0.44 (1/0)<br>2.01 (0/-1)<br>2.12 (-1/-2) | 0.61 (3/2)<br>0.79 (2/-1)<br>1.24 (-1/-2)<br>1.74 (-2/-3) | 0.02 (2/1)<br>0.82 (1/-2)<br>2.04 (-2/-3)<br>2.42 (-3/-4) | 1.18 (2/1)<br>1.45 (1/0)<br>1.94 (0/-1)<br>2.58 (-1/-2) | 0.11 (2/1)<br>0.39 (1/0)<br>1.96 (0/-1)<br>2.29 (-1/-2) |
| Defect | Bi$_{Ga}$ | Bi$_{As}$ | Ga$_{As}$ | v$_{Ga}$ | As$_{Ga}$ | v$_{As}$ | Bi$_{Ga}$+Bi$_{As}$ | Bi$_{Ga}$+Ga$_{As}$ | v$_{Ga}$+Bi$_{As}$ | v$_{Ga}$+v$_{As}$ | As$_{Ga}$+Bi$_{As}$ | As$_{Ga}$+Ga$_{As}$ |
| GaAsBi (eV) | 0.37 (2/1)<br>0.83 (1/0) | -0.10 (1/0) | -0.03 (1/0)<br>0.43 (0/-1)<br>0.73 (-1/-2) | 0.21 (0/-2)<br>0.69 (-2/-3) | 0.40 (2/1)<br>0.80 (1/0) | 0.20 (3/1)<br>0.82 (1/0)<br>0.93 (0/-1)<br>1.19 (-1/-2)<br>1.58 (-2/-3)<br>1.64 (-3/-4) | 0.58 (2/1)<br>0.96 (1/0) | 0.11 (1/0)<br>1.54 (0/-1) | 0.37 (3/-1)<br>0.69 (-1/-2)<br>1.15 (-2/-3) | 0.55 (1/-2)<br>1.37 (-2/-3) | 0.61 (2/1)<br>1.00 (1/0)<br>1.67 (0/-1) | 0.06 (1/0)<br>1.38 (0/-1) |
| Defect | Bi$_{Ga}$ | Bi$_{Sb}$ | Ga$_{Sb}$ | v$_{Ga}$ | Sb$_{Ga}$ | v$_{Sb}$ | Bi$_{Ga}$+Bi$_{Sb}$ | Bi$_{Ga}$+Ga$_{Sb}$ | v$_{Ga}$+Bi$_{Sb}$ | v$_{Ga}$+v$_{Sb}$ | Sb$_{Ga}$+Bi$_{Sb}$ | Sb$_{Ga}$+Ga$_{Sb}$ |
| GaSbBi (eV) | -0.06 (2/1)<br>0.31 (1/0) | | 0.15 (-1/-2) | 0.05 (-1/-2)<br>0.32 (-2/-3) | 0.23 (2/1)<br>0.54 (1/0) | 0.08 (1/-1)<br>0.45 (-1/-2)<br>0.65 (-2/-3)<br>0.77 (-3/-4) | -0.01 (2/1)<br>0.39 (1/0) | 0.79 (0/-1) | 0.17 (-1/-2)<br>0.44 (-2/-3) | 0.35 (0/-2)<br>0.38 (-2/-3)<br>0.67 (-3/-4) | 0.27 (2/1)<br>0.61 (1/0) | 0.89 (0/-1) |
| Defect | Bi$_{In}$ | Bi$_P$ | In$_P$ | v$_{In}$ | P$_{In}$ | v$_P$ | Bi$_{In}$+Bi$_P$* | Bi$_{In}$+In$_P$* | v$_{In}$+Bi$_P$* | v$_{In}$+v$_P$* | P$_{In}$+Bi$_P$* | P$_{In}$+In$_P$* |
| InPBi (eV) | 0.31 (2/1)<br>0.77 (1/0) | -0.14 (2/1)<br>0.12 (1/0) | -0.12 (4/3)<br>0.12 (3/2)<br>0.38 (2/1)<br>0.58 (1/0)<br>1.02 (0/-1)<br>1.27 (-1/-2) | 0.22 (0/-1)<br>0.53 (-1/-2)<br>0.72 (-2/-3) | 0.83 (2/1)<br>1.03 (1/0) | 0.54 (3/1)<br>1.15 (1/0)<br>1.24 (0/-1) | 0.53 (2/1)<br>0.93 (1/0) | 0.04 (4/2)<br>0.39 (2/1)<br>0.77 (1/0) | 0.24 (3/1)<br>0.65 (1/0)<br>1.06 (0/-2)<br>1.38 (-2/-3) | 0.24 (1/0)<br>0.73 (0/-2)<br>1.38 (-2/-3) | 0.72 (2/1)<br>1.10 (1/0)<br>1.43 (0/-1) | 0.29 (2/1)<br>0.69 (1/0)<br>1.40 (0/-1) |
| Defect | Bi$_{In}$ | Bi$_{As}$ | In$_{As}$ | v$_{In}$ | As$_{In}$ | v$_{As}$ | Bi$_{In}$+Bi$_{As}$* | Bi$_{In}$+In$_{As}$* | v$_{In}$+Bi$_{As}$* | v$_{In}$+v$_{As}$* | As$_{In}$+Bi$_{As}$* | As$_{In}$+In$_{As}$* |
| InAsBi (eV) | 0.12 (2/1) | 0.00 (1/0) | 0.11 (2/1)<br>0.30 (1/0) | 0.21 (-1/-2) | 0.21 (2/1) | 0.44 (3/1) | 0.34 (2/1) | 0.30 (2/1) | 0.19 (3/1)<br>0.42 (1/-1) | 0.19 (1/0) | 0.26 (2/1) | 0.22 (2/1) |
| Defect | Bi$_{In}$ | Bi$_{Sb}$ | In$_{Sb}$ | v$_{In}$ | Sb$_{In}$ | v$_{Sb}$ | Bi$_{In}$+Bi$_{Sb}$* | Bi$_{In}$+In$_{Sb}$* | v$_{In}$+Bi$_{Sb}$* | v$_{In}$+v$_{Sb}$* | Sb$_{In}$+Bi$_{Sb}$* | Sb$_{In}$+In$_{Sb}$* |
| InSbBi (eV) | 0.11 (1/0) | | 0.06 (0/-1) | 0.18 (-2/-3) | 0.06 (2/1) | 0.13 (3/-1) | -0.01 (2/1) | 0.25 (1/0) | | 0.07 (1/-2) | 0.03 (2/1) | 0.23 (1/0) |

TABLE III. Charge-state transition levels in reference to the valence band maximum. The levels printed in gray lay outside of the gap but fit within 10% of the gap. Numbers in parenthesis indicate the charge-state transition. Energy values are given in electronvolts (eV). Rows calculated with the MBA approach are indicated with an asterisk.

Second nearest neighbour pair defects

| Defect | $Bi_{Ga}$+$Bi_{Ga}$ | $Bi_P$+$Bi_P$ | $Ga_P$+$Bi_P$ | $Ga_P$+$Ga_P$ | $v_{Ga}$+$Bi_{Ga}$ | $v_{Ga}$+$v_{Ga}$ | $v_{Ga}$+$P_{Ga}$ | $P_{Ga}$+$Bi_{Ga}$ | $P_{Ga}$+$P_{Ga}$ | $v_P$+$Bi_P$ | $v_P$+$Ga_P$ | $v_P$+$v_P$ |
|---|---|---|---|---|---|---|---|---|---|---|---|---|
| GaPBi (eV) | 0.14 (4/3)<br>0.29 (3/2)<br>1.00 (2/1)<br>1.17 (1/0)<br>2.47 (0/-1) | 0.14 (2/1)<br>0.58 (1/0) | -0.02 (2/1)<br>0.44 (1/0)<br>1.29 (0/-1)<br>1.73 (-1/-2) | 0.00 (2/1)<br>0.24 (1/0)<br>0.81 (0/-1)<br>1.01 (-1/-2)<br>2.27 (-2/-3) | 0.60 (3/1)<br>1.13 (1/-1)<br>1.90 (-1/-2)<br>2.08 (-2/-3) | 0.44 (0/-1)<br>0.62 (-1/-2)<br>0.98 (-2/-3)<br>1.18 (-3/-4)<br>1.78 (-4/-5) | -0.18 (2/1)<br>0.27 (1/0)<br>0.46 (0/-1)<br>2.02 (-1/-2)<br>2.39 (-2/-3) | -0.22 (4/3)<br>0.30 (3/2)<br>1.21 (2/1)<br>1.52 (1/0)<br>1.60 (0/-2)<br>2.57 (-2/-3) | -0.10 (4/3)<br>0.08 (3/2)<br>1.43 (2/0)<br>2.23 (0/-3) | 0.20 (3/1)<br>1.12 (1/-1)<br>2.09 (-1/-2)<br>2.41 (-2/-3)<br>2.43 (-3/-4) | 0.10 (3/2)<br>0.44 (2/0)<br>1.05 (0/-1)<br>1.54 (-1/-3)<br>2.58 (-3/-4) | 0.09 (4/2)<br>0.74 (2/0)<br>1.28 (0/-1)<br>1.56 (-1/-2)<br>1.81 (-2/-3)<br>1.99 (-3/-4) |

| Defect | $Bi_{Ga}$+$Bi_{Ga}$ | $Bi_{As}$+$Bi_{As}$ | $Ga_{As}$+$Bi_{As}$ | $Ga_{As}$+$Ga_{As}$ | $v_{Ga}$+$Bi_{Ga}$ | $v_{Ga}$+$v_{Ga}$ | $v_{Ga}$+$As_{Ga}$ | $As_{Ga}$+$Bi_{Ga}$ | $As_{Ga}$+$As_{Ga}$ | $v_{As}$+$Bi_{As}$ | $v_{As}$+$Ga_{As}$ | $v_{As}$+$v_{As}$ |
|---|---|---|---|---|---|---|---|---|---|---|---|---|
| GaAsBi (eV) | -0.07 (4/3)<br>0.11 (3/2)<br>0.71 (2/1)<br>0.88 (1/0) | 0.17 (1/0) | 0.06 (1/0)<br>0.65 (0/-1)<br>1.05 (-1/-2) | -0.06 (1/0)<br>0.34 (0/-1)<br>0.55 (-1/-2)<br>1.47 (-2/-3) | 0.44 (3/-1)<br>1.30 (-1/-2)<br>1.62 (-2/-3) | 0.05 (0/-1)<br>0.23 (-1/-2)<br>0.57 (-2/-3)<br>0.78 (-3/-4)<br>1.41 (-4/-5) | -0.06 (1/0)<br>0.17 (0/-1)<br>1.24 (-1/-2)<br>1.63 (-2/-3) | 0.25 (3/2)<br>0.56 (2/1)<br>0.95 (1/0)<br>1.42 (0/-1)<br>1.65 (-1/-2) | 0.07 (3/2)<br>0.75 (2/1)<br>0.93 (1/0)<br>1.51 (0/-2) | 0.47 (1/-1)<br>1.10 (-1/-3) | 0.22 (2/0)<br>0.32 (0/-1)<br>1.02 (-1/-3) | 0.01 (4/2)<br>0.41 (2/0)<br>0.74 (0/-1)<br>0.96 (-1/-2)<br>1.25 (-2/-3)<br>1.50 (-3/-4) |

| Defect | $Bi_{Ga}$+$Bi_{Ga}$ | $Bi_{Sb}$+$Bi_{Sb}$ | $Ga_{Sb}$+$Bi_{Sb}$ | $Ga_{Sb}$+$Ga_{Sb}$ | $v_{Ga}$+$Bi_{Ga}$ | $v_{Ga}$+$v_{Ga}$ | $v_{Ga}$+$Sb_{Ga}$ | $Sb_{Ga}$+$Bi_{Ga}$ | $Sb_{Ga}$+$Sb_{Ga}$ | $v_{Sb}$+$Bi_{Sb}$ | $v_{Sb}$+$Ga_{Sb}$ | $v_{Sb}$+$v_{Sb}$ |
|---|---|---|---|---|---|---|---|---|---|---|---|---|
| GaSbBi (eV) | 0.13 (2/1)<br>0.38 (1/0) | | -0.07 (0/-1)<br>0.19 (-1/-2) | -0.02 (-1/-2)<br>0.36 (-2/-3)<br>0.69 (-3/-4) | -0.08 (3/0)<br>-0.04 (0/-1)<br>0.65 (-1/-2) | 0.16 (-2/-3)<br>0.41 (-3/-4)<br>0.84 (-4/-5) | -0.05 (0/-1)<br>0.74 (-1/-2) | 0.30 (2/1)<br>0.60 (1/-1) | 0.02 (3/2)<br>0.37 (2/1)<br>0.64 (1/0) | 0.07 (-1/-2)<br>0.34 (-2/-3) | -0.05 (2/-1)<br>0.07 (-1/-3)<br>0.32 (-3/-4)<br>0.56 (-4/-5) | -0.02 (0/-2)<br>0.06 (-2/-3)<br>0.43 (-3/-4)<br>0.57 (-4/-5) |

| Defect | $Bi_{In}$+$Bi_{In}$ | $Bi_P$+$Bi_P$ | $In_P$+$Bi_P$ | $In_P$+$In_P$ | $v_{In}$+$Bi_{In}$ | $v_{In}$+$v_{In}$ | $v_{In}$+$P_{In}$ | $P_{In}$+$Bi_{In}$ | $P_{In}$+$P_{In}$ | $v_P$+$Bi_P$ | $v_P$+$In_P$ | $v_P$+$v_P$ |
|---|---|---|---|---|---|---|---|---|---|---|---|---|
| InPBi (eV)* | 0.09 (3/2)<br>0.53 (2/1)<br>0.99 (1/0) | 0.27 (2/1)<br>0.74 (1/0) | -0.10 (4/2)<br>0.34 (2/1)<br>0.76 (1/0)<br>1.31 (0/-1) | -0.09 (3/2)<br>0.26 (2/1)<br>0.63 (1/0)<br>0.97 (0/-1)<br>1.41 (-1/-2) | 0.28 (3/1)<br>1.08 (1/-1)<br>1.22 (-1/-2) | -0.13 (0/-1)<br>0.22 (-1/-2)<br>0.54 (-2/-3)<br>0.86 (-3/-4)<br>1.27 (-4/-5) | 0.20 (0/-1)<br>1.29 (-1/-2) | 0.78 (2/1)<br>1.14 (1/0)<br>1.48 (0/-3) | 0.71 (2/1)<br>1.18 (1/0)<br>1.53 (0/-1) | 0.02 (2/1)<br>0.78 (1/-1) | 0.27 (2/1)<br>0.73 (1/0)<br>0.79 (0/-1)<br>1.20 (-1/-2) | 0.34 (2/1)<br>0.37 (1/0)<br>0.59 (0/-1)<br>1.06 (-1/-2)<br>1.42 (-2/-3) |

| Defect | $Bi_{In}$+$Bi_{In}$ | $Bi_{As}$+$Bi_{As}$ | $In_{As}$+$Bi_{As}$ | $In_{As}$+$In_{As}$ | $v_{In}$+$Bi_{In}$ | $v_{In}$+$v_{In}$ | $v_{In}$+$As_{In}$ | $As_{In}$+$Bi_{In}$ | $As_{In}$+$As_{In}$ | $v_{As}$+$Bi_{As}$ | $v_{As}$+$In_{As}$ | $v_{As}$+$v_{As}$ |
|---|---|---|---|---|---|---|---|---|---|---|---|---|
| InAsBi (eV)* | 0.01 (3/2)<br>0.43 (2/1) | 0.24 (2/1) | 0.23 (2/1) | 0.12 (2/1) | 0.40 (3/-1) | 0.06 (-1/-2)<br>0.35 (-2/-3) | 0.21 (0/-1) | 0.34 (2/1) | 0.26 (2/1) | 0.01 (2/1) | 0.11 (2/1) | 0.09 (2/1)<br>0.24 (1/-1) |

| Defect | $Bi_{In}$+$Bi_{In}$ | $Bi_{Sb}$+$Bi_{Sb}$ | $In_{Sb}$+$Bi_{Sb}$ | $In_{Sb}$+$In_{Sb}$ | $v_{In}$+$Bi_{In}$ | $v_{In}$+$v_{In}$ | $v_{In}$+$Sb_{In}$ | $Sb_{In}$+$Bi_{In}$ | $Sb_{In}$+$Sb_{In}$ | $v_{Sb}$+$Bi_{Sb}$ | $v_{Sb}$+$In_{Sb}$ | $v_{Sb}$+$v_{Sb}$ |
|---|---|---|---|---|---|---|---|---|---|---|---|---|
| InSbBi (eV)* | 0.10 (2/1) | 0.26 (1/0) | -0.01 (1/0) | 0.19 (0/-1) | 0.10 (0/-1) | 0.04 (-2/-3) | 0.11 (0/-1) | 0.19 (2/1) | 0.23 (2/1) | | 0.03 (-1/-2) | |

add-ons/files. The code used to postprocess the raw data is available on figshare [81]: https://doi.org/10.6084/m9.figshare.12478700

### D. Competing Interest

The authors declare no competing interest.